\documentclass[twocolumn, superscriptaddress, showpacs, preprintnumbers, amsmath, amssymb]{revtex4}
\usepackage{graphicx}

\begin{document}

\title{Asymmetric crystallization during cooling and heating in model glass-forming systems}

\author{Minglei Wang}
\affiliation{Department of Mechanical Engineering and Materials Science, Yale University, New Haven, Connecticut, 06520, USA}
\affiliation{Center for Research on Interface Structures and Phenomena, Yale University, New Haven, Connecticut, 06520, USA}

\author{Kai Zhang}
\affiliation{Department of Mechanical Engineering and Materials Science, Yale University, New Haven, Connecticut, 06520, USA}
\affiliation{Center for Research on Interface Structures and Phenomena, Yale University, New Haven, Connecticut, 06520, USA}

\author{Zhusong Li}
\affiliation{Department of Physics and Benjamin Levich Institute, The City College of the City University of New York, New York, New York, 10031, USA}

\author{Yanhui Liu}
\affiliation{Department of Mechanical Engineering and Materials Science, Yale University, New Haven, Connecticut, 06520, USA}
\affiliation{Center for Research on Interface Structures and Phenomena, Yale University, New Haven, Connecticut, 06520, USA}

\author{Jan Schroers}
\affiliation{Department of Mechanical Engineering and Materials Science, Yale University, New Haven, Connecticut, 06520, USA}
\affiliation{Center for Research on Interface Structures and Phenomena, Yale University, New Haven, Connecticut, 06520, USA}

\author{Mark D. Shattuck}
\affiliation{Department of Mechanical Engineering and Materials Science, Yale University, New Haven, Connecticut, 06520, USA}
\affiliation{Department of Physics and Benjamin Levich Institute, The City College of the City University of New York, New York, New York, 10031, USA}

\author{Corey S. O'Hern}
\affiliation{Department of Mechanical Engineering and Materials Science, Yale University, New Haven, Connecticut, 06520, USA}
\affiliation{Center for Research on Interface Structures and Phenomena, Yale University, New Haven, Connecticut, 06520, USA}
\affiliation{Department of Physics, Yale University, New Haven, Connecticut, 06520, USA}
\affiliation{Department of Applied Physics, Yale University, New Haven, Connecticut, 06520, USA}

\date{\today}

\begin{abstract}
We perform molecular dynamics (MD) simulations of the crystallization
process in binary Lennard-Jones systems during heating and cooling to
investigate atomic-scale crystallization kinetics in glass-forming
materials. For the cooling protocol, we prepared equilibrated liquids
above the liquidus temperature $T_l$ and cooled each sample to zero
temperature at rate $R_c$.  For the heating protocol, we first cooled
equilibrated liquids to zero temperature at rate $R_p$ and then heated
the samples to temperature $T > T_l$ at rate $R_h$.  We measured the
critical heating and cooling rates $R_h^*$ and $R_c^*$, below which
the systems begin to form a substantial fraction of crystalline
clusters during the heating and cooling protocols.  We show that
$R_h^* > R_c^*$, and that the asymmetry ratio $R_h^*/R_c^*$ includes
an intrinsic contribution that increases with the glass-forming
ability (GFA) of the system and a preparation-rate dependent
contribution that increases strongly as $R_p \rightarrow R_c^*$ from
above. We also show that the predictions from classical nucleation
theory (CNT) can qualitatively describe the dependence of the
asymmetry ratio on the GFA and preparation rate $R_p$ from the MD
simulations and results for the asymmetry ratio measured in Zr- and
Au-based bulk metallic glasses (BMG).  This work emphasizes the need
for and benefits of an improved understanding of crystallization
processes in BMGs and other glass-forming systems.
\end{abstract}

\pacs{64.70.pe,
 64.70.Q-,
 61.43.Fs,
 61.66.Dk 
} 

\maketitle

\section{introduction}

Crystallization, during which a material transforms from a dense,
amorphous liquid to a crystalline solid, occurs via the nucleation and
subsequent growth of small crystalline
domains~\cite{Frenkel:1939}. Crystallization in metals has been
intensely studied over the past several decades with the goal of
developing the ability to tune the microstructure to optimize the
mechanical properties of metal
alloys~\cite{Doherty:1997,Asta:2009,Boettinger:2000}.  However,
in-situ observation of crystallization in metallic melts is limited
due to the rapid crystallization kinetics of
metals~\cite{Hollandmoritz:1993, Schroers:2000, Shen:2009}.

In contrast, bulk metallic glasses (BMGs), which are amorphous metal
alloys, can be supercooled to temperatures below the solidus
temperature $T_s$ and persist in a dense, amorphous liquid state over
more than $12$ orders of magnitude in time scales or
viscosity~\cite{Busch:2000}. Deep supercooling of BMGs provides the
ability to study crystallization on time scales that are accessible to
experiments~\cite{Kelton:1991, Busch:2007, Ehmler:1998, Masuhr:1999}.

These prior experimental studies have uncovered fundamental questions
concerning crystallization kinetics in BMGs.  For example, when a BMG
in the glass state is heated to a temperature $T_f < T_s$ in the
supercooled liquid region, crystallization is much faster than
crystallization that occurs when the metastable melt is cooled to the same
temperature $T_f$~\cite{Perepezko:2004, Perepezko:2003}. Asymmetries
in the crystallization time scales upon heating versus cooling of up
to two orders of magnitude have been reported in
experiments~\cite{Jan:1999, Pogatscher:2014}.  The asymmetry impacts
industrial applications of BMGs because rapid crystallization upon
heating limits the thermoplastic forming processing time window for
BMGs~\cite{Jan:2010, Jan:2013, Johnson:2011,Pitt:2011}.

Recent studies have suggested that the asymmetry in the
crystallization time scales originates from the temperature dependence
of the the nucleation and growth rates~\cite{Jan:1999}, {\it i.e.}
that the nucleation rate is maximal at a temperature below that at
which the growth rate is maximal. According to this argument,
crystallization upon heating is faster because of the growth of the
nascent crystal nuclei that formed during the thermal quench to the
glass.  In contrast, crystallization is slower upon cooling since
crystal nuclei are not able to form at high temperatures in the melt.
However, there has been no direct visualization of the crystallization
process in BMGs, and it is not yet understood why the asymmetry varies
from one BMG to another~\cite{Lu:2002} and how sensitively the
asymmetry depends on the cooling rate $R_p$ used to prepare the
glass. An improved, predictive understanding of the crystallization
process in BMGs will aid the design of new BMG-forming alloys with
small crystallization asymmetry ratios and large thermoplastic
processing time windows.

\begin{figure*}[t]
\begin{center}
$\begin{array}{cccccc}
\includegraphics[width = 1.0 in]{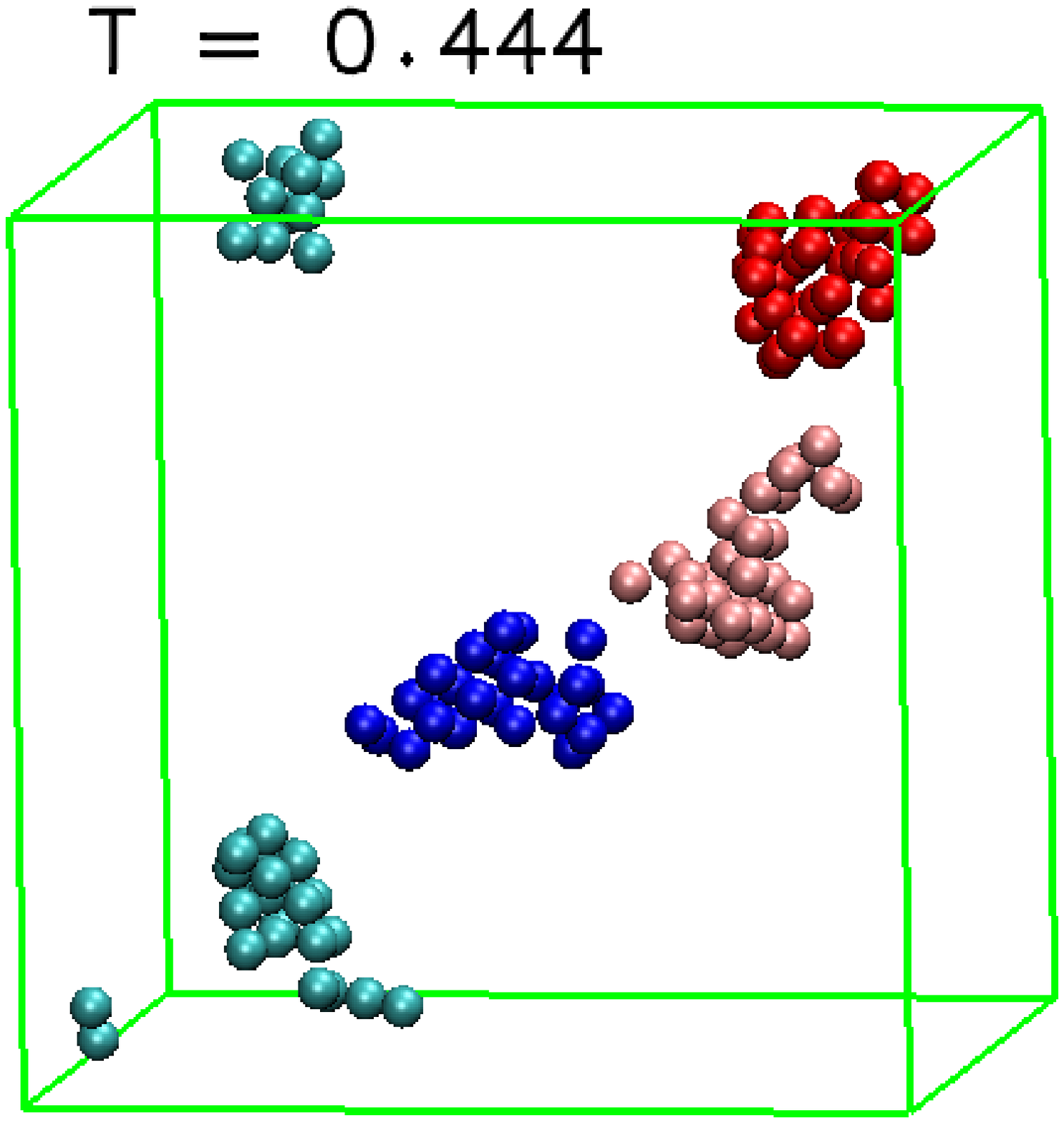} & \includegraphics[width = 1.0 in]{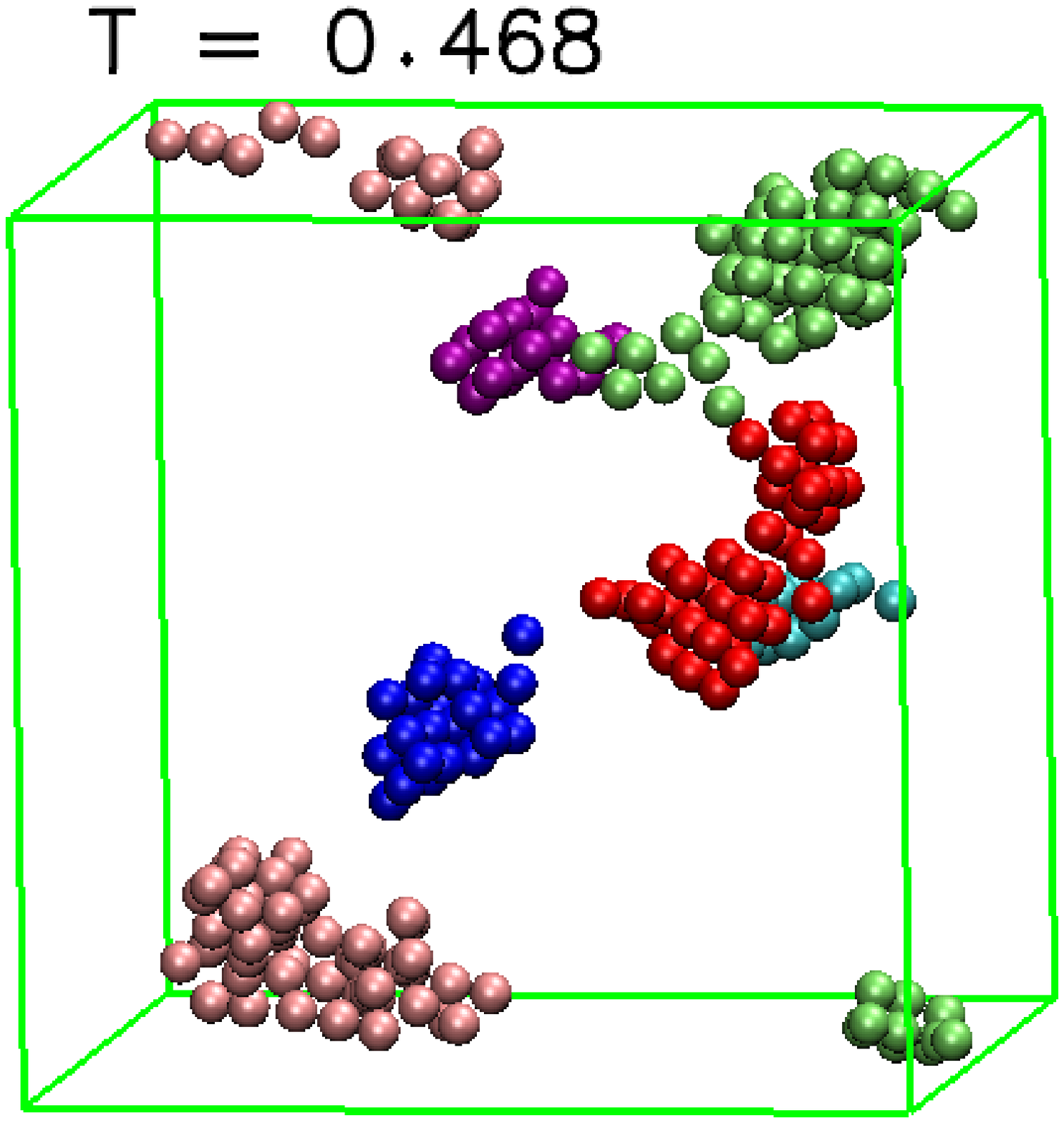} & \includegraphics[width = 1.0 in]{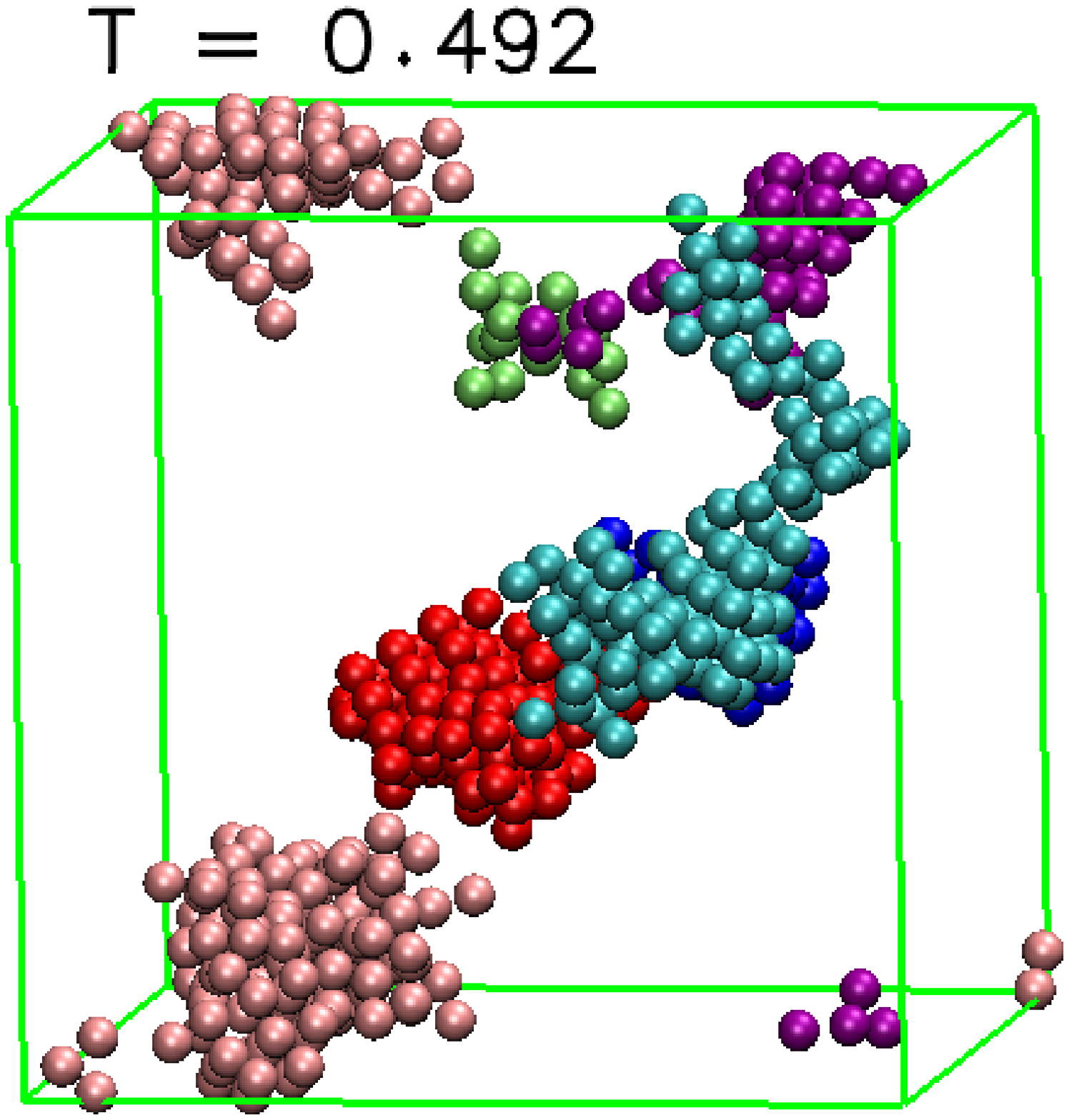} & \includegraphics[width = 1.0 in]{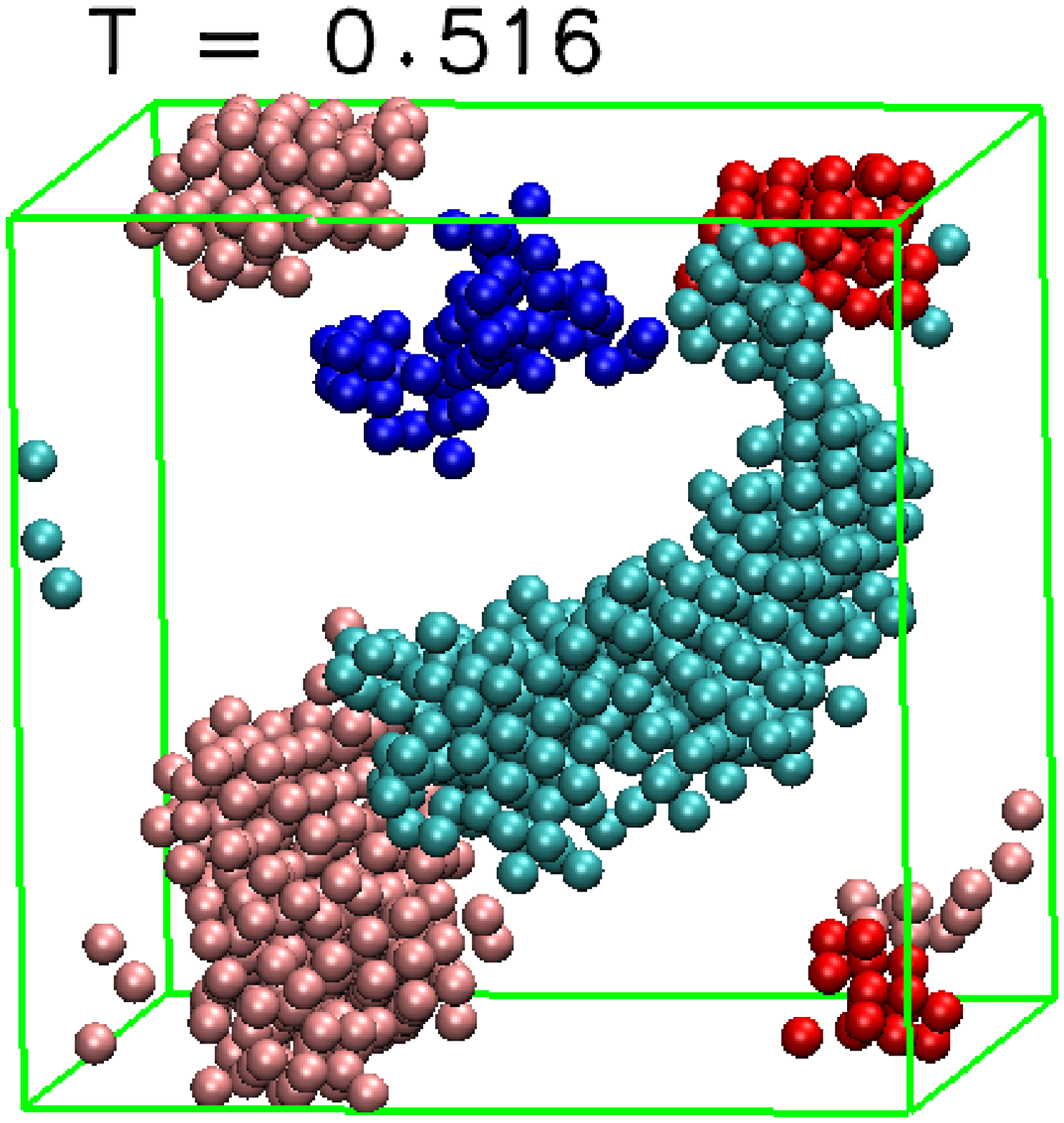} & \includegraphics[width = 1.0 in]{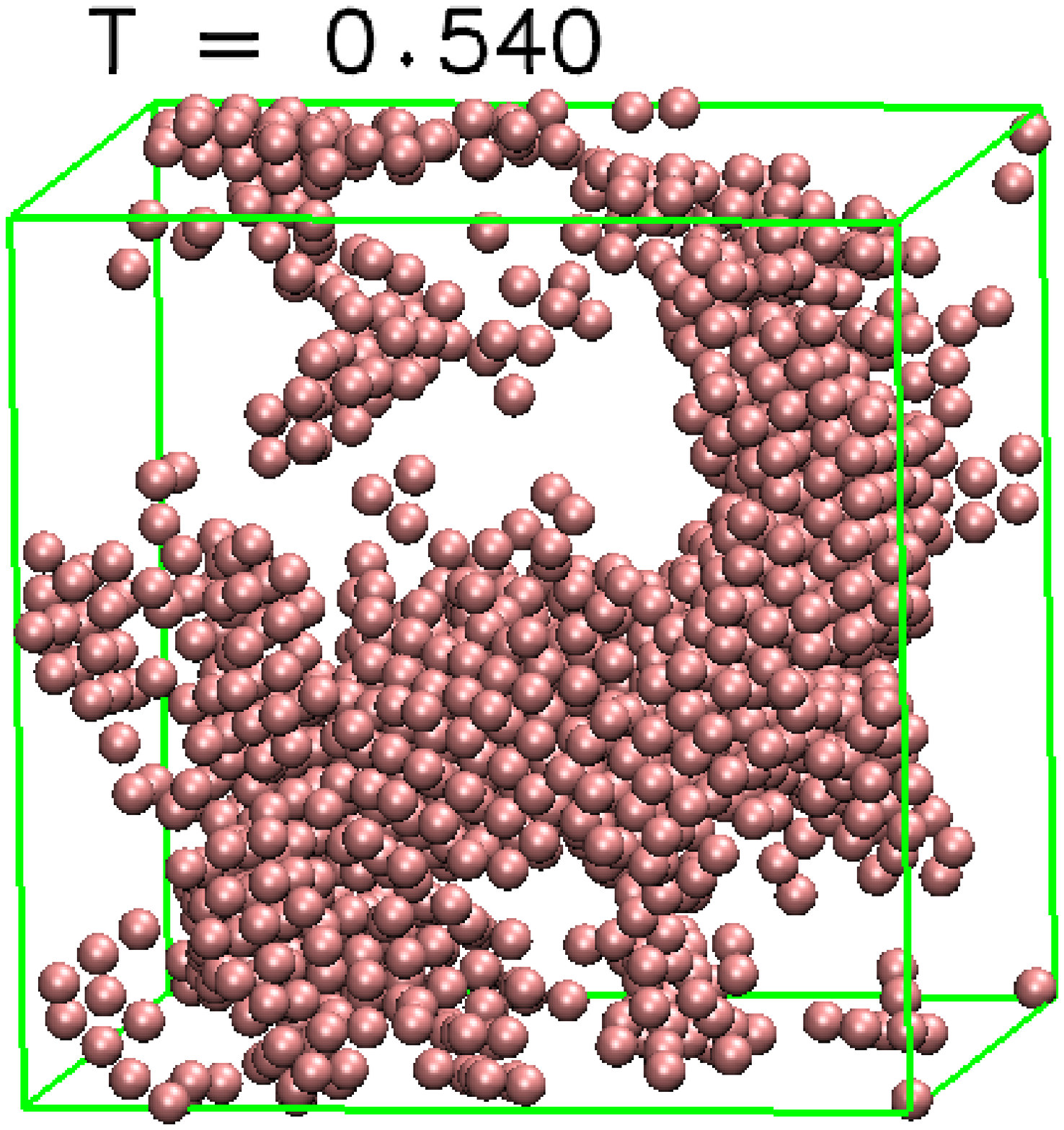} & \includegraphics[width = 1.2 in]{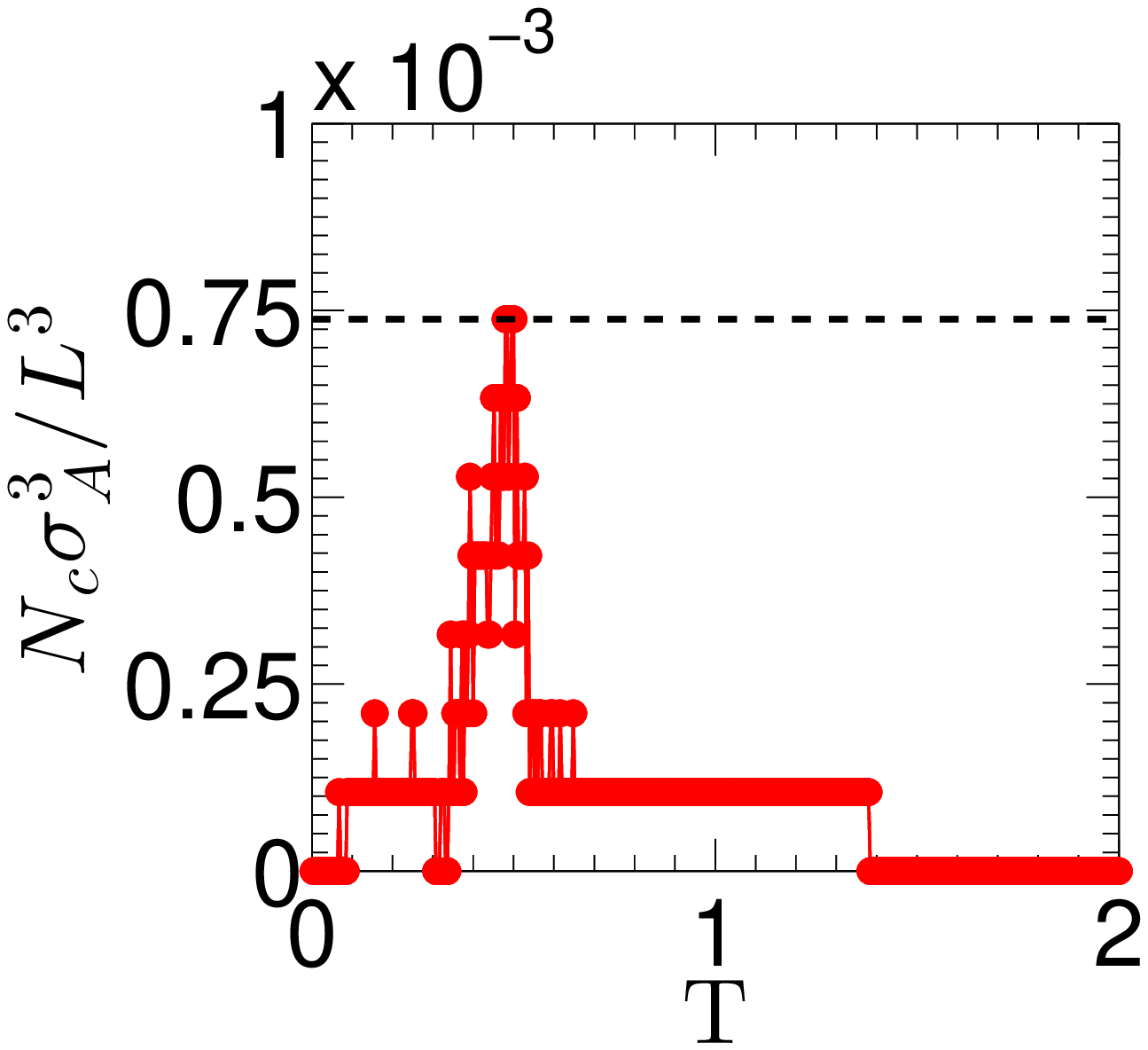} \\
\includegraphics[width = 1.0 in]{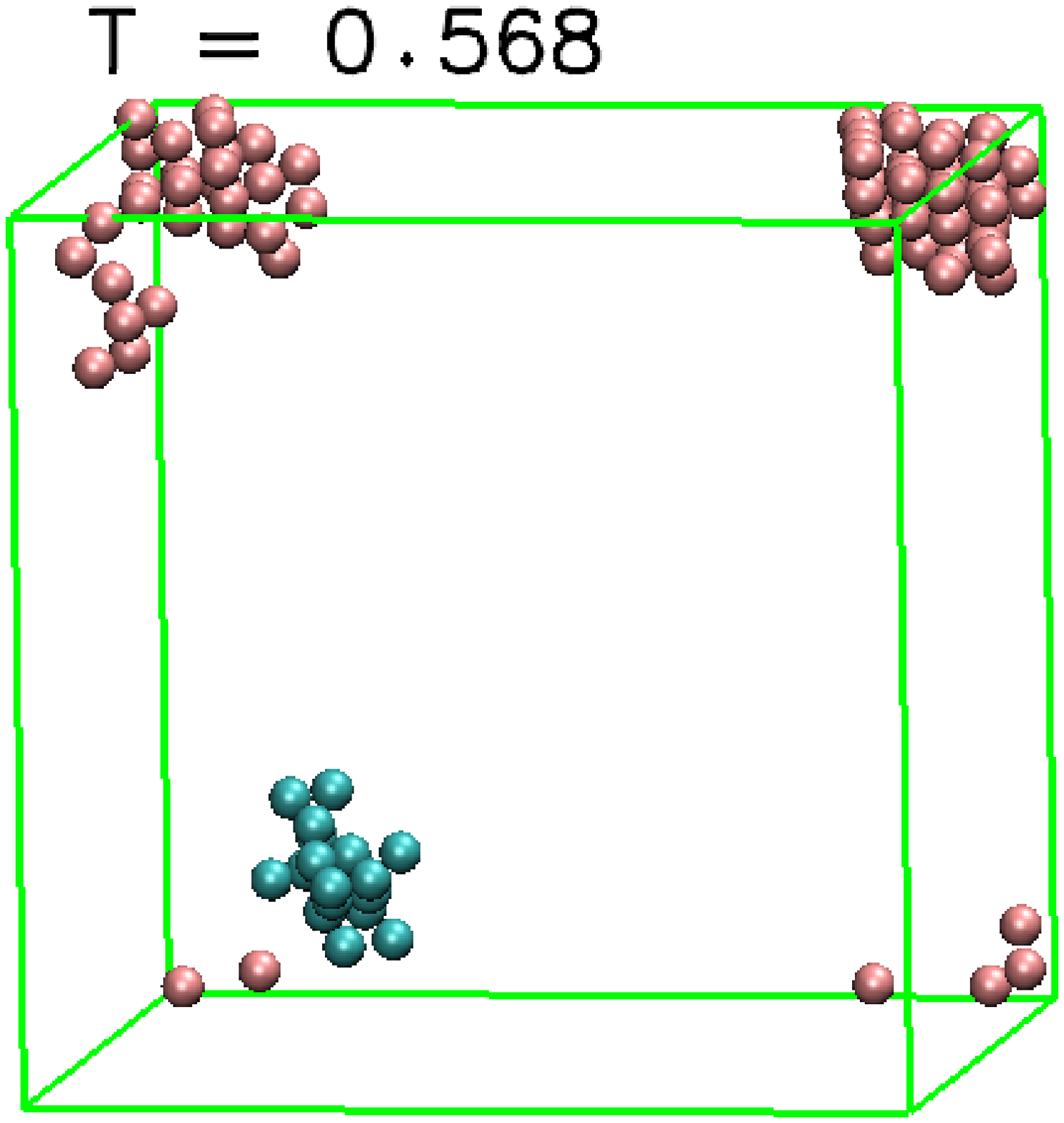} & \includegraphics[width = 1.0 in]{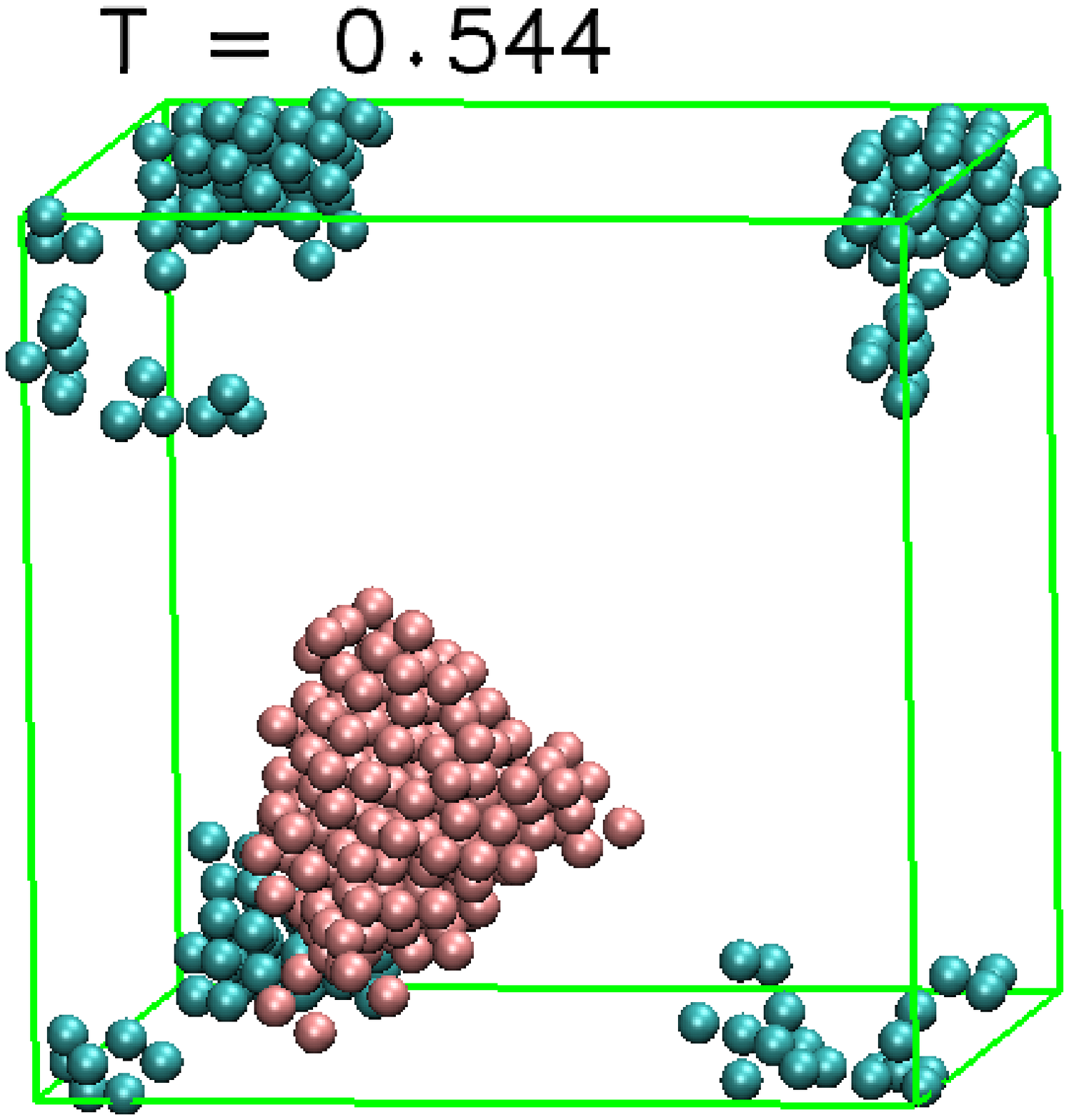} & \includegraphics[width = 1.0 in]{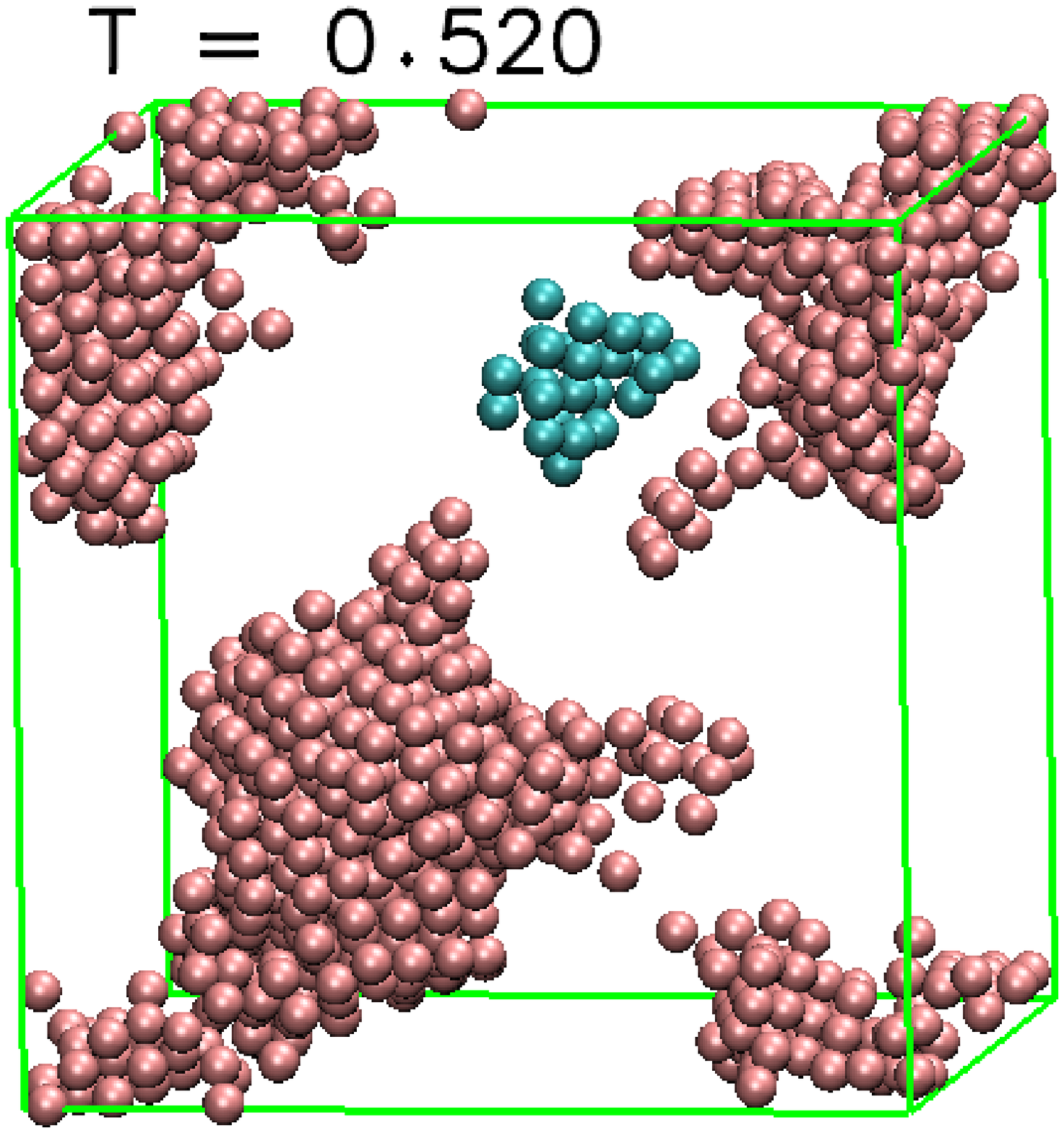} & \includegraphics[width = 1.0 in]{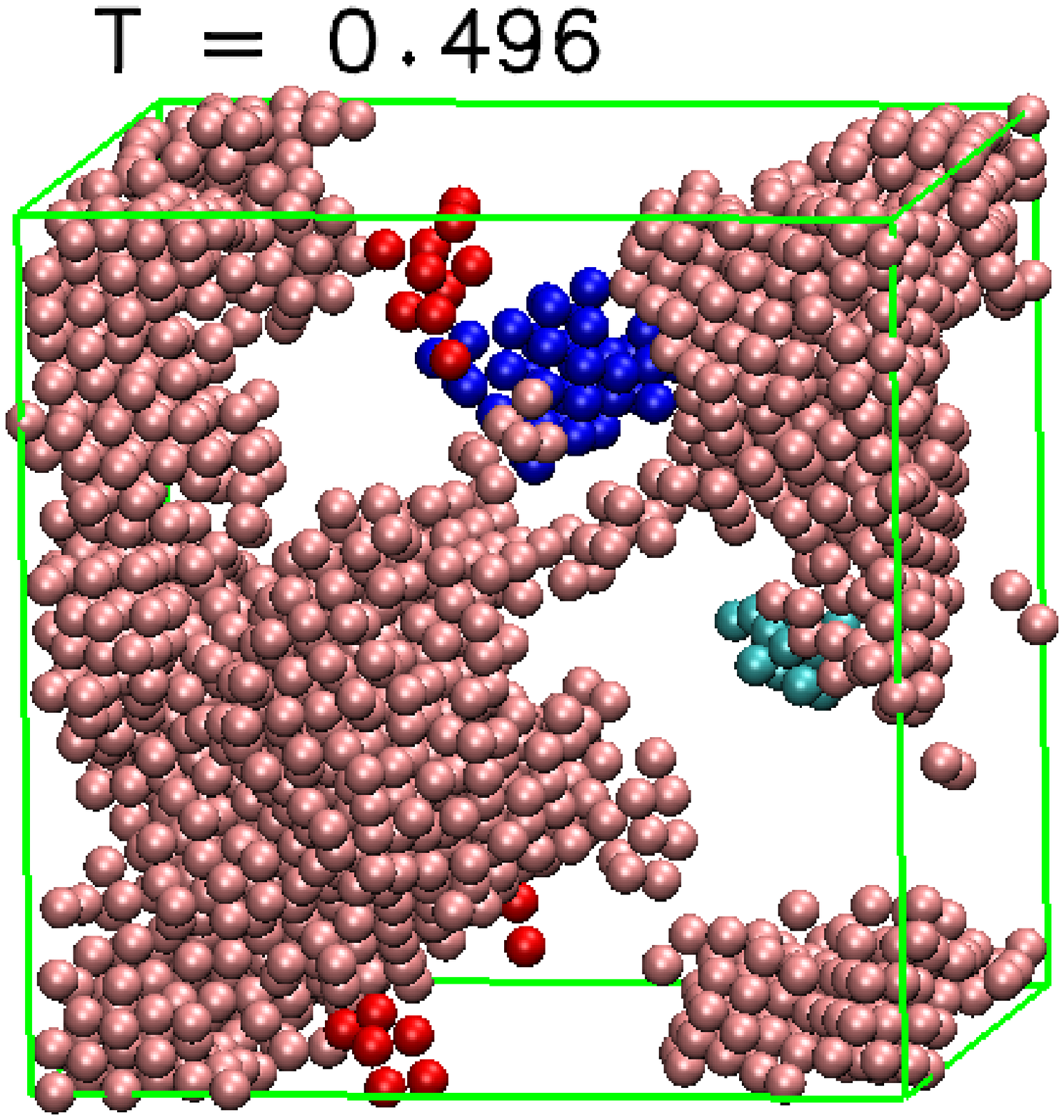} & \includegraphics[width = 1.0 in]{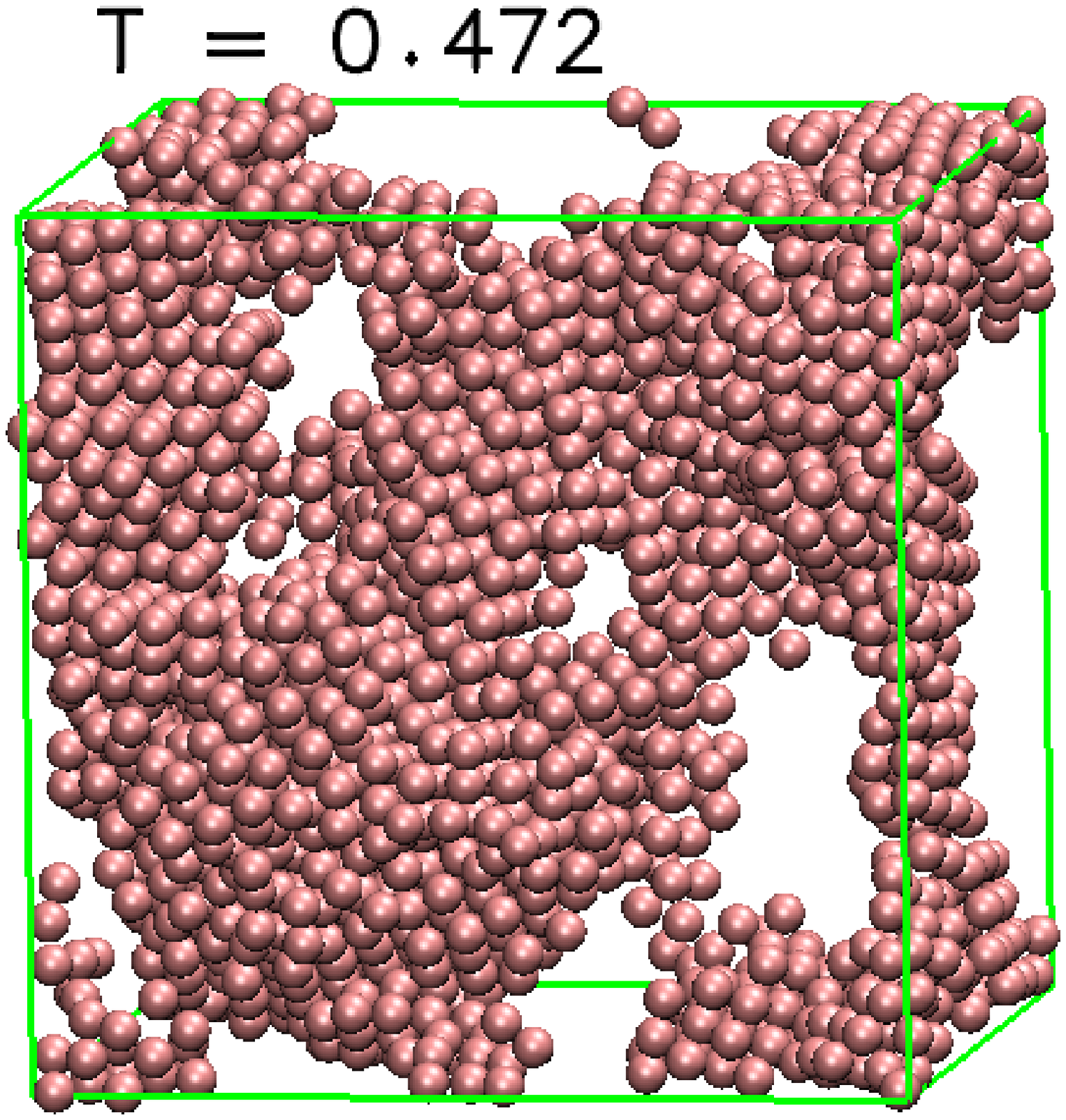} & \includegraphics[width = 1.2 in]{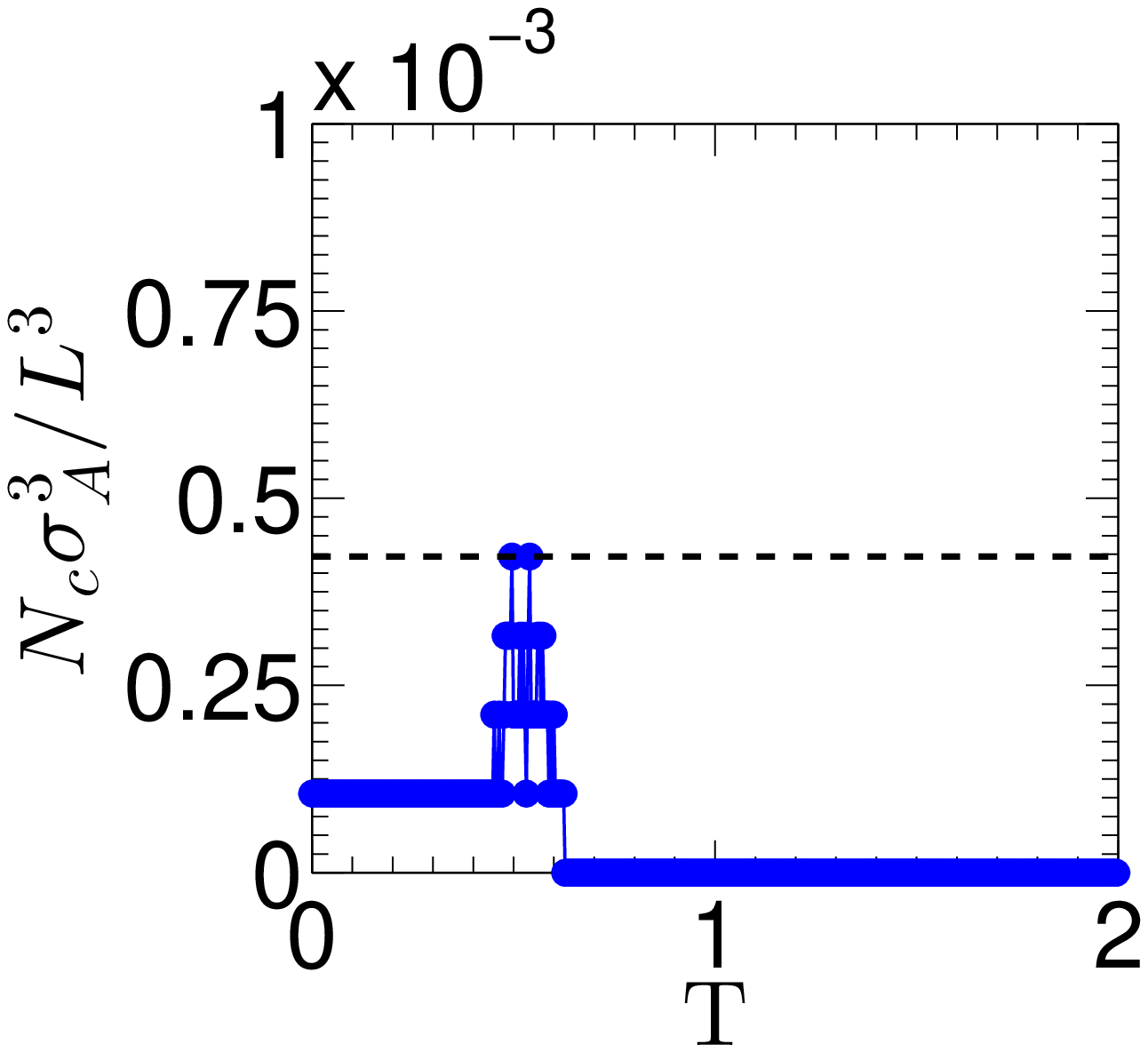} \\
\end{array}$
\end{center}
\caption{(Color online) Snapshots of the nucleation and growth of
crystal clusters at several temperatures $T$ as a monodisperse Lennard-Jones 
system is heated
from zero temperature to $T_f=2.0$ at a rate $R_h < R_h^*$ (top row)
and cooled from initial temperature $T_i=2.0$ to zero temperature at a
rate $R_c < R_c^*$ (bottom row). Distinct, disconnected crystal nuclei
are shaded different colors.  The far right panel indicates the number
of clusters $N_{c}$ normalized by $L^3/\sigma_A^3$ of
as a function of temperature during a
typical heating (top) and cooling (bottom) trajectory. The maximum number 
of clusters $N_c^{\rm max}$ is indicated by the horizontal dashed line.}
\label{fig:snapshot}
\end{figure*}

We employ molecular dynamics (MD) simulations of
bidisperse spheres interacting via Lennard-Jones
potentials~\cite{Kob:1999,Johnson:1998,Zhang:2013} to visualize
directly the crystallization process upon heating and cooling in model 
metallic glass-forming systems. We
perform thermal quenches of the system from a high temperature $T_i$
in the equilibrated liquid regime to a glass at $T_f = 0$ and vary the
cooling rate $R_c$ by several orders of magnitude.  For cooling rates
below the critical cooling rate $R_c < R_c^*$, the system begins to
crystallize, whereas for $R_c > R_c^*$, the system remains amorphous.
We also performed MD simulations in which we heat the zero-temperature
glassy states (prepared at cooling rate $R_p > R_c^*$) through the
supercooled liquid regime to $T_f = T_i$ over a range of heating rates
$R_h$. For heating rates $R_h < R_h^*$, the system begins to
crystallize, whereas for $R_h > R_h^*$, it remains amorphous. We also
find that the critical heating rate has an intrinsic contribution
$R_h^*(\infty)$ and an $R_p$-dependent contribution
$R_h^*(R_p)-R_h^*(\infty)$ that increases with decreasing $R_p$. We
measured the asymmetry ratio $R_h^*/R_c^*$ as a function of the
glass-forming ability (GFA) and $R_p$ for several binary Lennard-Jones
mixtures and find that $R_h^*/R_c^* > 1$ and the ratio grows with
increasing GFA and decreasing $R_p$.  We show that these results are
consistent with predictions from classical nucleation theory (CNT) that the
maximal growth rate occurs at a higher temperature than the maximal
nucleation rate and that the separation between the nucleation and
growth peaks increases with the GFA.  Further,
CNT is able to qualitatively recapitulate the
dependence of the asymmetry ratio on the GFA as measured through
$R_c^*$ for both our MD simulations and recent experiments on BMGs as
well as on $R_p$ for the MD simulations~\cite{Nishiyama:1999}.

The remainder of the manuscript is organized into three sections:
Sec.~\ref{sec:method}: Methods, Sec.~\ref{sec:results}: Results, and
Sec.~\ref{sec:discussion}: Conclusion.  In Sec.~\ref{sec:method}, we
describe the MD simulations of binary Lennard-Jones mixtures, the
computational methods to detect and structurally characterize crystal
nuclei, and measurements of the critical cooling and heating rates,
$R_c^*$ and $R_h^*$.  In Sec.~\ref{sec:results}, we show results from
MD simulations for the time-temperature transformation
diagram~\cite{ttt} and the asymmetry ratio $R_h^*/R_c^*$ as a function
of the glass-forming ability as measured by the critical cooling rate
$R_c^*$ and the cooling rate used to prepare the zero-temperature
glasses $R_p$.  We also compare our results for the asymmetry ratio to
experimental measurements of the ratio for two BMGs and to predictions
of the ratio from classical nucleation theory.  In
Sec.~\ref{sec:discussion}, we briefly summarize our results and put 
forward our conclusions.

\section{Methods}
\label{sec:method}

We performed MD simulations of binary Lennard-Jones (LJ) mixtures of
$N=N_A+N_B$ spheres with mass $m$ at constant volume $V=L^3$ in a
cubic simulation box with side length $L$ and periodic boundary conditions.  We
studied mixtures with $N_A=N_B$ and diameter ratio $\alpha =
\sigma_B/\sigma_A < 1$. We employed the LJ pairwise interaction
potential between spheres $i$ and $j$:
\begin{equation}
u(r_{ij}) = 4\epsilon [(\sigma_{ij} / r_{ij})^{12} - (\sigma_{ij} / r_{ij})^6],
\end{equation}
where $r_{ij}$ is their center-to-center separation, $\epsilon$ is the
depth of the minimum in the potential energy $u(r_{ij})$, $\sigma_{ij}
= (\sigma_i + \sigma_j)/2$, and $u(r_{ij})$ has been truncated and
shifted so that the potential energy and force vanish for separations
$r_{ij} \ge 3.5\sigma_{ij}$~\cite{Allen}. We varied the system volume
$V$ to fix the packing fraction $\phi = \pi \sigma_A^3(N_A + \alpha^3
N_B)/6V = 0.5236$~\cite{Barroso:2002} at each diameter ratio
$\alpha$. For most simulations, we considered $N=1372$ spheres, but we
also studied $N=4000$ and $8788$ to assess finite-size effects. Below,
energy, length, time, and temperature scales are expressed in units of
$\epsilon$, $\sigma_A$, $\sigma_{A} \sqrt{m /\epsilon}$, and
$\epsilon/k_B$, respectively, where the Boltzmann constant $k_B$ has
been set to be unity.

\subsection{Cooling and Heating Protocols}
\label{protocol}

For each particle diameter ratio, which yield different glass-forming 
abilities, we performed MD simulations to cool
metastable liquids to zero temperature and heat zero-temperature
glasses into the metastable liquid regime to measure $R_c^*$ and
$R_h^*$ at which the systems begin to crystallize. To measure $R_c^*$,
we first equilibrate the system at high temperature $T_i = 2.0$ using
a Gaussian constraint thermostat~\cite{Allen}. We then cool the system
by decreasing the temperature linearly at rate $R_c$ from $T_i$ to
$T_f = 0$:
\begin{equation}
\label{linear}
T(t) = T_i - R_c t.
\end{equation}

To measure the critical heating rate $R_h^*(R_p)$ at finite rate
$R_p$, we first prepare the systems in a glass state by cooling them from
the high temperature liquid state to zero-temperature at rate $R_p >
R_c^*$. To measure the intrinsic critical heating rate
$R_h^*(\infty)$, we quench the systems infinitely fast to zero
temperature using conjugate gradient energy minimization.  For both
cases, we heat the zero-temperature glasses using a linear ramp
\begin{equation}
\label{heating}
T(t) = R_h t
\end{equation}
until $T_f = 2.0$. For both heating and cooling protocols, we carried
out $N_{\rm tot} = 1000$ independent trajectories and averaged the results.

\subsection{Identification of Crystal Nuclei}

To detect the onset of crystallization in our
simulations~\cite{Jan:1999}, we differentiate `crystal-like' versus
`liquid-like' particles based on the value of the area-weighted bond
orientational order parameter for each
particle~\cite{Jungblut:2011,Wolde:1996}. We define the complex-valued
bond orientational order parameter for particle $i$:
\begin{equation}
q_{lm}(i) = \frac{\sum^{N_b}_{j = 1} A_{ij} Y_{lm}(\theta( {\vec r}_{ij}),\phi({\vec r}_{ij})) }{\sum^{N_b}_{j = 1} A_{ij}},
\end{equation}
where $Y_{lm}(\theta({\vec r}_{ij}), \phi({\vec r}_{ij}))$ is the
spherical harmonic of degree $l$ and order $m$, $\theta({\vec
r}_{ij})$ and $\phi({\vec r}_{ij})$ are the polar and azimuthal angles
for the vector ${\vec r}_{ij}$, $j=1,\ldots,N_b$ gives the index of
the Voronoi neighbors of particle $i$, and $A_{ij}$ is the area of the
face of the Voronoi polyhedron common to particles $i$ and $j$. The
correlation coefficient~\cite{Jungblut:2011} between the bond
orientational order parameters $q_{lm}(i)$ and $q_{lm}(j)$, where
particle $j$ is a Voronoi neighbor of $i$,
\begin{equation}
S_{ij} = \frac{\sum^{6}_{m = -6}q_{6m}(i)q^{\ast}_{6m}(j)}{\left(\sum^{6}_{m = -6} \mid q_{6m}(i)\mid^2\right)^{1/2}\left(\sum^{6}_{m = -6} \mid q_{6m}(j)\mid^2\right)^{1/2}}
\end{equation}
is sensitive to face-centered-cubic (FCC) order.  When $S_{ij} > 0.7$,
$i$ and $j$ are considered `connected'. If particle $i$ has more than
$10$ connected Voronoi neighbors, it is defined as `crystal-like'. The
ratio $N_{cr}/N$ gives the fraction of crystal-like particles in a
given configuration.  In addition, we also define a crystal cluster as
the set of crystal-like particles that possess mutual Voronoi
neighbors.  Distinct crystal clusters that nucleate and grow upon
heating and cooling are shown in Fig.~\ref{fig:snapshot}.

\begin{figure}
\begin{center}
$\begin{array}{c}
\includegraphics[width = 3.5 in]{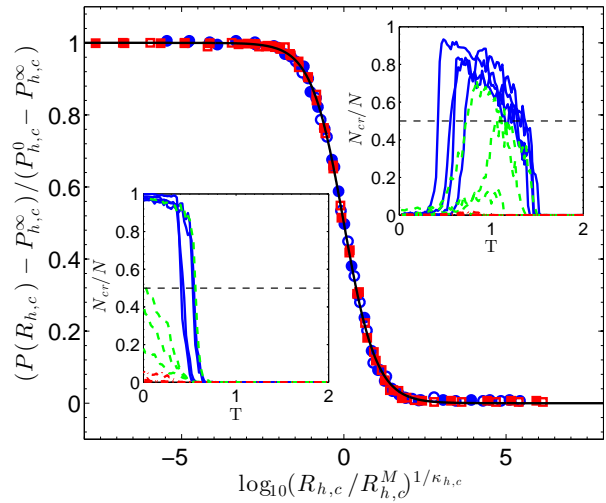}
\end{array}$
\end{center}
\caption{Shifted and normalized probability for crystallization
$(P(R_{h,c})-P^{\infty}_{h,c})/(P^0_{h,c}-P^{\infty}_{h,c})$ versus
the scaled heating or cooling rate $\log_{10}
(R_{h,c}/R^M_{h,c})^{1/\kappa_{h,c}}$.  Circles (squares) indicate
data for cooling (heating) for diameter ratios $\alpha=1.0$ (filled
symbols) and $0.97$ (open symbols). The insets show the fraction of
crystal-like particles $N_{cr}/N$ as a function of temperature $T$
during cooling (lower left) and heating (upper right) for $12$
configurations with $\alpha=1.0$. The four solid, dashed, and
dot-dashed curves in each inset correspond to cooling and heating
trajectories with rates slower than $R^*_{h,c}$, near $R^*_{h,c}$, and
faster than $R^*_{h,c}$, respectively. Trajectories for which
$N_{cr}/N $ exceeds $0.5$ (above the horizontal dashed line) are
considered to have crystallized during the heating or cooling
protocol.}
\label{fig:coolheat}
\end{figure}

\subsection{Probability for Crystallization}

For each diameter ratio and rate, we measure the probability for
crystallization $P(R_{h,c}) = N_X / N_{\rm tot}$, where $N_X$ is the
number of trajectories that crystallized with $N_{cr}/N > 0.5$ during
the heating or cooling protocol and $N_{\rm tot}$ is the total number
of trajectories ({\it cf.} insets to Fig.~\ref{fig:coolheat}).  We
find that the data for $P(R_{h,c})$ collapses onto a sigmoidal scaling
function as shown in Fig.~\ref{fig:coolheat}:
\begin{equation}
\label{pcrystal}
\frac{(P(R_{h,c}) - P^{\infty}_{h,c})}{P^{0}_{h,c} - P^{\infty}_{h,c}} = 
\frac{1}{2}\left[1 - \tanh\left(\log_{10}\left(\frac{R_{h,c}}{R^M_{h,c}}\right)^{1/\kappa_{h,c}}\right)\right],
\end{equation}
where $P^{\infty}_{h,c}$ is the probability for crystallization in the
limit of infinitely fast rates $R_{h,c} \rightarrow \infty$,
$P^{0}_{h,c}$ is the probability for crystallization in the $R_{h,c}
\rightarrow 0$ limit, $R^M_{h,c}$ is the rate at which $P(R_{h,c}) =
(P^{0}_{h,c} + P^{\infty}_{h,c})/2$, and $\kappa_{h,c}$ is the
stretching factor. We find that $\kappa_c \approx 0.25$ and $\kappa_h
\approx 0.2$ for $\alpha = 1.0$, and these factors increase by only a
few percent over the range in $\alpha$ that we consider.  We define
the critical heating and cooling rates $R^*_h$ and $R^*_c$ by the
rates at which $P(R_{h,c})=0.5$, {\it i.e.}
\begin{equation}
R^*_{h,c} = R^M_{h,c} 10^{\kappa_{h,c} \tanh^{-1} \left[\frac{P^{0}_{h,c} + P^{\infty}_{h,c}-1}{P^{0}_{h,c} - P^{\infty}_{h,c}} \right]}.
\end{equation}
As shown in the insets to Fig.~\ref{fig:coolheat}, for $R_{h,c} \ll
R^*_{h,c}$ most of the configurations crystallize during heating or
cooling.  In contrast, for $R_{h,c} \gg R^*_{h,c}$, none of the
configurations crystallize.

\begin{figure}
\includegraphics[width = 3.5 in]{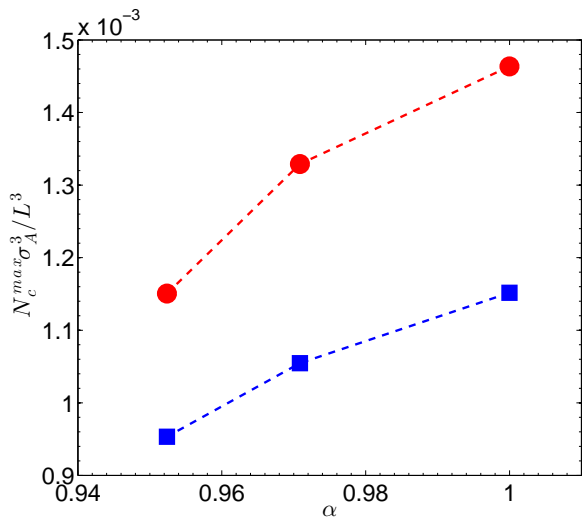}
\caption{Maximum value $N_c^{\rm max}$ of the number of crystal
clusters $N_c(T)$ normalized by $L^3/\sigma_A^3$ (averaged over $1000$ 
trajectories) during the cooling (squares) and
heating (circles) protocols at rates $R_c \approx 0.5 R_c^*$ and $R_h
\approx 0.5 R_h^*$ for LJ mixtures with diameter ratios $\alpha =
1.0$, $0.97$, and $0.95$. For all systems, the maximum number of
crystal clusters is larger for the heating protocol compared to that
for the cooling protocol and $N_c^{\rm max}$ decreases with increasing
glass-forming ability (decreasing $\alpha$).}
\label{fig:nClusters}
\end{figure}

\section{Results}
\label{sec:results}

An advantage of MD simulations is that they can provide atomic-level
structural details of the crystallization dynamics that are often
difficult to obtain in experiments. In Fig.~\ref{fig:snapshot}, we
visualize the nucleation and growth of clusters of crystal-like
particles during the heating and cooling simulations.  In both cases,
the number of clusters reaches a maximum near $T \approx
0.5$. In Fig.~\ref{fig:nClusters}, we show the maximum number
of clusters $N_c^{\rm max}$ (normalized by $L^3/\sigma_A^3$) that form
during the heating and cooling protocols.  We find that more crystal
clusters form during the heating protocol compared to the cooling
protocol for all particle diameter ratios studied, which is supported
by the measured time-temperature-transformation (TTT) diagram. In
addition, we will show below that the asymmetry ratio $R_h^*/R_c^* >
1$, and that the ratio grows with increasing GFA (increasing diameter
ratio) and decreasing $R_p$.  We find that CNT can qualitatively
describe the dependence of the asymmetry ratio on the GFA, as measured
by the critical cooling rate $R_c^*$, for both our MD simulations and
recent experiments on BMGs, as well as on the preparation cooling rate
$R_p$ for the MD simulations.

\subsection{Intrinsic Asymmetry Ratio}
\label{intrinsic}

\begin{figure}
\includegraphics[width = 3.5 in]{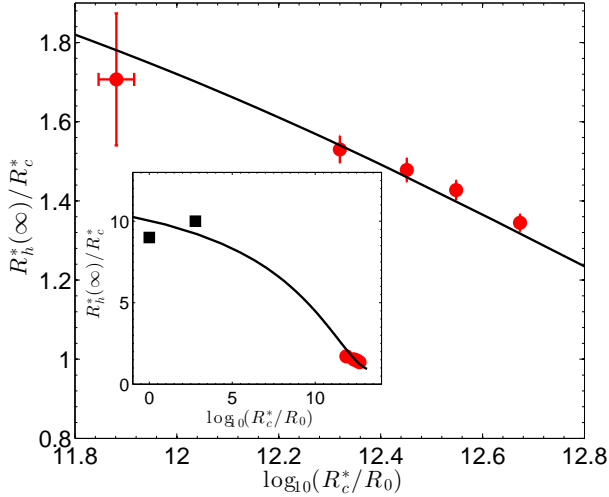}
\caption{Intrinsic asymmetry ratio $R^*_h(\infty)/R_c^*$ versus the
critical cooling rate $R^*_c$ (for diameter ratios $\alpha = 1.0$,
$0.97$, $0.96$, $0.95$, and $0.93$) normalized by $R_0 = 1 K/s$ on a
logarithmic scale. The inset shows the intrinsic asymmetry ratio
versus $\log_{10} R^*_c/R_0$ on an expanded scale.  The filled circles
indicate data from the MD simulations and filled squares indicate data
from experiments on Zr- and Au-based
BMGs~\cite{Jan:1999,Pogatscher:2014}.  The prediction
(Eq.~\ref{eq:Rc}) from classical nucleation theory (solid line) with
$A' = (8\pi A D_0^4)/3 a^3 = 0.5$ (in units of $\epsilon^2/(m^2
\sigma_A^4)$), $\Sigma = 0.26$, and $Q_{eff}= 2.6$ interpolates
between the MD simulation data at high $R_c^*$ and experimental data
from BMGs at low $R_c^*$.}
\label{fig:RcEffect}
\end{figure}

The critical heating and cooling rates can be obtained by fitting the
probability for crystallization $P(R_{h,c})$ as a function of $R_h$ or
$R_c$ to the sigmoidal form in Eq.~\ref{pcrystal}.  We first
investigate the minimum value for the asymmetry ratio
$R^*_h(\infty)/R^*_c$, which is obtained by taking the $R_p
\rightarrow \infty$ limit. (The asymmetry ratio $R^*_h(R_p)/R_c^*$ for
finite preparation rates $R_p$ will be considered in Sec.~\ref{rp}.)
In Fig.~\ref{fig:RcEffect}, we plot $R^*_h(\infty)/R^*_c$ versus
$R^*_c$ (for diameter ratios $\alpha=1.0$, $0.97$, $0.96$, $0.95$, and
$0.93$). We find that $R^*_h(\infty) > R^*_c$ for all systems studied,
which is consistent with classical nucleation theory (CNT). As shown
in Fig.~\ref{fig:snapshot}, more crystal nuclei form during the
heating protocol than during the cooling protocol. In addition, CNT
predicts that the growth rates for crystal nuclei are larger during heating
compared to cooling. In Sec.~\ref{cnt_prediction}, we will show that
both factors contribute to an increased probability for
crystallization during heating.

In Fig.~\ref{fig:RcEffect}, we also show that the asymmetry ratio 
$R_h^*(\infty)/R^*_c$ increases as the critical cooling rate $R_c^*$ 
decreases, or equivalently as the glass-forming ability increases. In 
the MD simulations, we were able to show a correlation between 
the asymmetry ratio and the critical cooling rate over roughly an order 
of magnitude in $R_c^*$. In Sec.~\ref{cnt_prediction}, we introduce 
a model that describes qualitatively this dependence of the asymmetry ratio on 
$R_c^*$.  

\subsection{Classical Nucleation Theory Prediction for the 
Asymmetry Ratio}
\label{cnt_prediction}

In classical nucleation theory (CNT), the formation of crystals is a
two-step process: 1) fluctuations in the size of crystal nuclei that
allow them to reach the critical radius $r^*$ and 2) growth of
post-critical nuclei with $r>r^*$. To form a critical nucleus, the
system must overcome a nucleation free energy barrier:
\begin{equation}
\Delta G^{\ast} = \frac{16\pi}{3} \frac{\Sigma^3}{\Delta G^2},
\end{equation}
where $\Delta G$ is the bulk Gibbs free energy difference per volume
(in units of $\epsilon/\sigma_A^3$) and $\Sigma$ is the surface tension
between the solid and liquid phases (in units of
$\epsilon/\sigma_A^2$). We assume that $\Delta G = c (T_m -
T)$~\cite{Sherif}, where $T_m$ is melting temperature, $T_m - T$ is
the degree of undercooling, and $c \sim L_v/T_m$ is a dimensionless
parameter that characterizes the thermodynamic drive to crystallize
and will be used to tune the GFA of the system (where $L_v$ is the
latent heat of fusion). Within CNT, the rate of formation of critical
nuclei (nucleation rate) is given by:
\begin{equation}
\label{nucleation}
I = AD_0\exp\left(-\frac{Q_{eff}}{T}\right) \exp\left(-\frac{\Delta G^{\ast}}{T}\right),
\end{equation} 
where $A$ is an $O(1)$ constant with units $\sigma_A^{-5}$, $D_0$ is
the atomic diffusivity with units $\sigma_A \sqrt{\epsilon/m}$, and
$Q_{eff}$ is an effective activation energy for the diffusivity (with
units of $\epsilon$). After the nucleation free energy barrier $\Delta
G^{\ast}$ has been overcome and crystal nuclei reach $r\ge r^*$, the
growth rate of crystal nuclei is given by
\begin{equation}
U = \frac{D_0}{a} \exp\left(-\frac{Q_{eff}}{T}\right) \left[1 - \exp\left(-\frac{\Delta GV}{T}\right)\right],
\end{equation}
where $a$ the characteristic interatomic spacing. 

\begin{figure}
\includegraphics[width = 3.5 in]{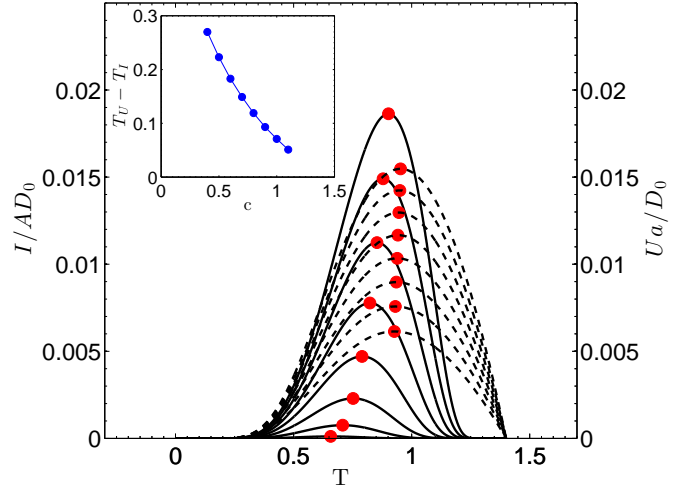}
\caption{The nucleation $I/AD_0$ (solid lines; left axis) and growth
$Ua/D_0$ (dashed lines; right axis) rates as a function of temperature
$T$ for increasing values of the glass-forming ability (GFA) $c=1.2$,
$1.1$, $1.0$, $0.9$, $0.8$, $0.7$, $0.6$, and $0.5$ (from top to
bottom) that span the range of diameter ratios from $\alpha=1.0$ to
$0.93$. The filled circles indicate the maximum rates ($I^*$ and
$U^*$) for each GFA.  As the GFA increases, $I^*$ and $U^*$ decrease
and the difference $T_U - T_I$ between the temperatures at which the
maxima in $U(T)$ and $I(T)$ occur increases (inset).}
\label{fig:IUschematic}
\end{figure}

In Fig.~\ref{fig:IUschematic}, we plot the nucleation $I/AD_0$ and
growth rates $Ua/D_0$ with $Q_{eff} = 2.6$ and $T_m \approx
1.40$ from MD simulations of binary LJ systems~\cite{Meier:2004},
$\Sigma = 0.26$, which is typical for
BMGs~\cite{Jan:1999}, while varying the GFA parameter from $c=1.2$ to
$0.5$ (corresponding to diameter ratios from $\alpha=1.0$ to $0.93$.)
Both $I(T)$ and $U(T)$ are peaked with maxima $I^*$ and $U^*$ at
temperatures $T_I$ and $T_U$.  In Fig.~\ref{fig:IUschematic}, we show
that as the GFA increases, $I^*$ and $U^*$, as well as $T_I$ and $T_U$
decrease.  However, $T_I$ decreases faster than $T_U$, so that the
separation between the peaks, $T_U - T_I$, increases with GFA.

To determine the critical heating and cooling rates, $R_h^*$ and
$R_c^*$, we must calculate the fraction of the samples $N_X$ that
crystallize and the probability for crystallizing
$P(R_{h,c})=N_X/N_{\rm tot}$, where $N_{\rm tot}$ is the total number
of samples, upon heating and cooling.  Within classical nucleation
theory, the probability to crystallize upon cooling from $T_i$ to
$T_f$ is given by~\cite{Inoue}:
\begin{equation}
P(R_c) = \frac{4\pi}{3R_c^4} \int^{T_f}_{T_i} I(T')\left[\int^{T_f}_{T'} U(T'')dT''\right]^3 dT'.
\label{eq:X}
\end{equation}
We assume that $T_i$ is above the liquidus temperature $T_l$, and
$T_f$ is below the glass transition temperature $T_g$, where the time
required to form crystal nuclei diverges. We can rearrange
Eq.~\ref{eq:X} to solve for the critical cooling rate at which
$P(R_c^*)=0.5$:
\begin{equation}
\begin{split}
(R^*_{c})^4 &  = \frac{8\pi}{3} \int^{T_f}_{T_i} I(T') \left[\int^{T_f}_{T'} U(T'')dT''\right]^3 dT' \\
& = A' \int^{T_f}_{T_i} dT' \exp\left(-\frac{Q_{eff}}{T'}\right) \exp\left(-\frac{\Delta G^{\ast}}{T'}\right) \\
& \left[\int^{T_f}_{T'} \exp\left(-\frac{Q_{eff}}{T''}\right) \left[1 - \exp\left(-\frac{\Delta G V}{T''}\right)\right] dT''\right]^3,
\label{eq:Rc}
\end{split}
\end{equation}
where $A' = (8\pi A D^4_0)/(3 a^3)$ and we assumed that $A$,
$D_0$, and $a$ are independent of temperature. A similar expression
for the intrinsic critical heating rate $R_h^*(\infty)$ can be
obtained by reversing the bounds of integration in Eq.~\ref{eq:Rc}.

In Fig.~\ref{fig:RcEffect}, we plot the intrinsic asymmetry ratio
$R_h^*(\infty)/R_c^*$ predicted from Eq.~\ref{eq:Rc} versus the
critical cooling rate $R_c^*$ after choosing the best value $A' =
0.5$ that interpolates between the MD
simulation data at high $R_c^*$ and experimental data from BMGs at low
$R_c^*$. We find that CNT qualitatively captures the
increase in the asymmetry ratio with increasing GFA over a wide range
of critical cooling rates from $1 K/s$ (experiments on BMGs) to
$10^{12} K/s$ (MD simulations of binary LJ systems). A comparison of
Figs.~\ref{fig:RcEffect} and~\ref{fig:IUschematic} reveals that
the increase in the intrinsic asymmetry ratio is caused by the 
separation of the peaks in the growth and nucleation rates $U(T)$
and $I(T)$ that occurs as the GFA increases. Thus, we predict an enhanced
value for $T_U - T_I$ in experiments on BMGs since the critical
cooling rate in experiments is orders of magnitude smaller than in the
MD simulations.

\begin{figure}[t]
\begin{center}
\includegraphics[width = 3.5 in]{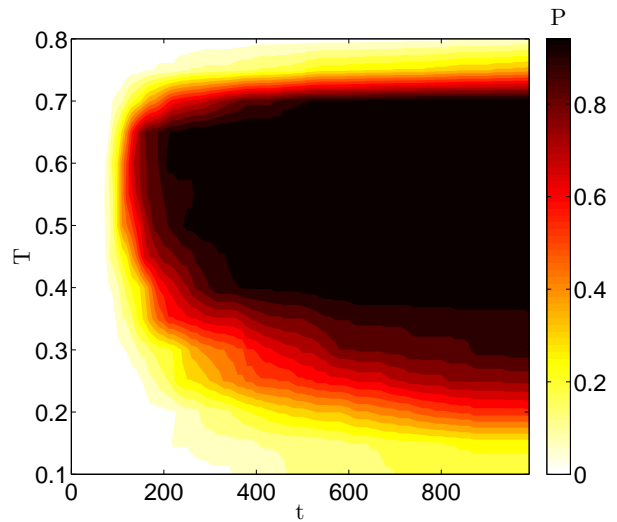}
\end{center}
\caption{The time-temperature-transformation (TTT) diagram during
cooling is visualized by plotting the probability to crystallize $P$
(increasing from light to dark) from $96$ samples as a function of
temperature $T$ and waiting time $t$ for LJ systems with diameter
ratio $\alpha = 1.0$. A sample is considered crystalline if the number
of crystal-like particles satisfies $N_{cr}/N > 0.5$. The initial
states are dense liquids equilibrated at $T = 2.0$. Each
initial state is cooled (at rate $R_c \gg R_c^*$) to temperature $T <
T_l$, where $T_l \approx 1.4$ is the liquidus temperature,
and then run at fixed $T$ for a time $t$.}
\label{fig:TTT}
\end{figure}

The fact that $R_h^*(\infty)>R_c^*$ is also reflected in the asymmetry
of the ``nose'' of the time-temperature-transformation (TTT)
diagram. In Fig~\ref{fig:TTT}, we show the probability $P$ that the
system has crystallized at a given temperature $T$ after a waiting
time $t$ for monodisperse LJ systems. We find that $T_{\min}\sim
0.5$-$0.6$ is the temperature at which the waiting time for
crystallization is minimized and that the time to crystallize is
in general longer for $T < T_{\rm min}$ compared to $T > T_{\rm min}$. Because
crystallization on average occurs at a higher temperature during
heating and a lower temperature during cooling, the asymmetry in the TTT
diagram indicates that slower rates are required to crystallize during
cooling than during heating, {\it i.e.}  $R_c^* < R_h^*$.

\subsection{Asymmetry Ratio for Finite $R_p$}
\label{rp}

\begin{figure}
\includegraphics[width = 3.5 in]{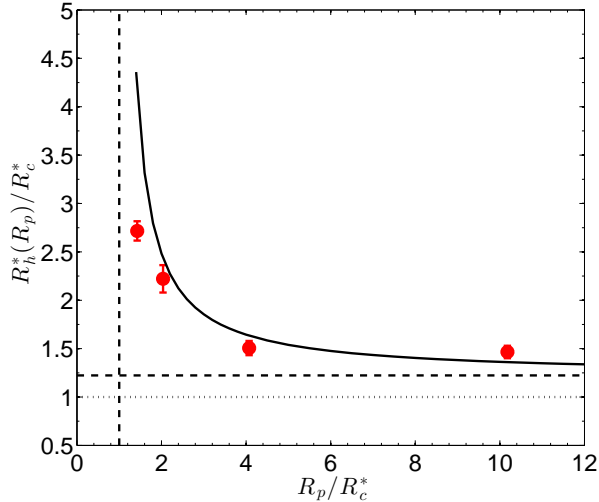}
\caption{Asymmetry ratio $R_h^*(R_p)/R_c^*$ plotted versus the
preparation cooling rate $R_p$ normalized by the critical cooling rate
$R_c^*$ from MD simulations with $\alpha = 1.0$ (filled circles) and
the prediction from CNT (solid line) with the same
parameters used for the fit in Fig.~\ref{fig:RcEffect} and the GFA
parameter set to $c=1.2$. The vertical dashed line indicates $R_p
=R_c^*$.  The horizontal dashed lines $R_h^*(R_p)/R_c^* = 1.18$ and $1$
indicate the plateau value in the $R_p \gg R_c^*$ limit and $R_h^* =
R_c^*$, respectively.  The gap between the horizontal dashed and dotted
lines give the magnitude of the intrinsic asymmetry ratio for this 
particular GFA ({\it cf.} Fig.~\ref{fig:RcEffect}).}
\label{fig:RpEffect}
\end{figure}

In Sec.~\ref{cnt_prediction}, we assumed that the initial samples
({\it i.e.} the zero-temperature glasses) for the heating protocol
were prepared in the $R_p \rightarrow \infty$ limit and, thus were
purely amorphous. How does the asymmetry ratio $R_h^*(R_p)/R_c^*$
depend on $R_p$ when the preparation cooling rate $R_p$ is finite and
partial crystalline order can occur in the samples?  In this section,
we show results for the asymmetry ratio $R_h^*(R_p)/R_c^*$ for
monodisperse systems using a protocol where the samples are quenched
from equilibrated liquid states to zero temperature at a finite rate
$R_p$ and then heated to temperature $T_f$ at rate $R_h$.  (See
Sec.~\ref{protocol}.)  Note that when $R_p/R_c^* \approx 1$, some of
the samples crystallize during the cooling preparation, yet these
samples are still included in the calculation of the probability
$P(R_h^*(R_p))$ to crystallize.  In Fig.~\ref{fig:RpEffect}, we show
the results for the asymmetry ratio $R_h^*(R_p)/R_c^*$ from MD
simulations. We find that $R_h^*(R_p)/R_c^*$ grows rapidly as $R_p$
approaches $R_c^*$ from above and reaches a plateau value of $\sim 1.2$ in
the limit $R_p/R_c^* \gg 1$.

The critical heating rate $R_h^*(R_p)$ at finite $R_p$ can also be
calculated from CNT using an expression similar to Eq.~\ref{eq:Rc}
with an additional term that accounts for cooling the equilibrated
liquid samples to zero temperature at a finite rate.  In
Fig.~\ref{fig:RpEffect}, we show that the asymmetry ratio
$R_h^*(R_p)/R_c^*$ predicted using CNT agrees qualitatively with that
from the MD simulations. The number of crystal nuclei that form during
the quench increases with decreasing $R_p$, which causes
$R_h^*(R_p)/R_c^*$ to diverge as $R_p \rightarrow R_c^*$. The
predicted {\it intrinsic} contribution to the asymmetry ratio for $R_p
\sim R_c^*$ is small, and $R_h^*(R_p)/R_c^*$ is dominated by the
preparation protocol.  In contrast, the asymmetry ratio
$R^*_h(R_p)/R_c^* \approx 1.2$ is dominated by the intrinsic
contribution in the $R_p \gg R_c^*$ limit.  As shown in
Fig.~\ref{fig:RcEffect}, the size of the intrinsic contribution to the
asymmetry ratio can be tuned by varying the GFA, which controls the
separation between the peaks in the nucleation $I(T)$ and growth $U(T)$
rates.
 
\section{Conclusion}
\label{sec:discussion}

We performed MD simulations of binary Lennard-Jones systems to model
the crystallization process during heating and cooling protocols in
metallic glasses. We focused on measurements of the ratio of the
critical heating $R_h^*$ and cooling $R_c^*$ rates, below which
crystallization occurs during the heating and cooling trajectories. We
find: 1) $R_h^* > R_c^*$ for all systems studied, 2) the asymmetry
ratio $R_h^*/R_c^*$ grows with increasing glass-forming ability (GFA),
and 3) the critical heating rate $R_h^*(R_p)$ has an intrinsic
contribution $R_h^*(\infty)$ and protocol-dependent contribution
$R_h^*(R_p)-R_h^*(\infty)$ that increases with decreasing cooling
rates $R_p$ used to prepare the initial samples at zero
temperature. We show that these results are consistent with the
prediction from classical nucleation theory that the maximal growth
rate occurs at a higher temperature than the maximal nucleation rate
and that the separation between the peaks in nucleation $I(T)$ and
growth $U(T)$ rates increases with the GFA.  Predictions from CNT are
able to qualitatively capture the dependence of the asymmetry ratio on
the GFA as measured through $R_c^*$ for both our MD simulations and
recent experiments on BMGs as well as on $R_p$ for the MD simulations.
Thus, our simulations have addressed how the thermal processing
history affects crystallization, which strongly influences the
thermoplastic formability of metallic glasses.

\begin{acknowledgments}
The authors acknowledge primary financial support from the NSF MRSEC
DMR-1119826. We also acknowledge support from the Kavli Institute
for Theoretical Physics (through NSF Grant No. PHY-1125915), where
some of this work was performed. This work also benefited from the
facilities and staff of the Yale University Faculty of Arts and
Sciences High Performance Computing Center and the NSF (Grant
No. CNS-0821132) that in part funded acquisition of the computational
facilities.
\end{acknowledgments}

\end{document}